\documentclass[twocolumn,epj]{webofc}

\usepackage[varg]{txfonts}   %
\usepackage{subcaption}
\usepackage{hyperref}
\usepackage{placeins}      %
\usepackage[sort&compress]{natbib}

\graphicspath{{Figs/}{./}}
\def\orcid#1{\kern .08em\href{https://orcid.org/#1}{\includegraphics[keepaspectratio,width=0.6em]{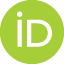}}}

\begin{document}
\title{UHECR Signatures and Sources}

\author{
    \firstname{Daniele} \lastname{Fargion}    
    \inst{1}       
    \fnsep\thanks{\email{daniele.fargion@fondazione.uniroma1.it}}
    \inst{\orcid{0000-0003-3146-3932}}
\and
    \firstname{Pier Giorgio} \lastname{De Sanctis Lucentini}
    \inst{2}
    \inst{\orcid{0000-0001-7503-2064}}
\and
    \firstname{Maxim Y.} \lastname{Khlopov}
    \inst{3,4}
    \inst{\orcid{0000-0002-1653-6964}}
}

\institute{
    Rome University “La Sapienza” and MIFP, Rome, Italy. 
\and
    National University of Oil and Gas «Gubkin University», Moscow, Russia.
\and
    National Research Nuclear University "MEPHI“, Moscow, Russia.
\and 
    Institute of Physics, Southern Federal University, Rostov on Don, Russia.
}

\abstract{%
We discuss recent results on the clustering, composition and distribution of Ultra-High Energy Cosmic Rays (UHECR) in the sky; from the energy of several tens of EeV in the  dipole anisotropy, up to the highest energy of a few narrow clusters, those of Hot Spots. 
Following the early UHECR composition records  deviations from proton we noted  that the UHECR events above 40 EeV can be made not just by any light or heavy nuclei, but mainly by the lightest ones as  He,D, Li,Be.
The remarkable Virgo absence and the few localized nearby extragalactic sources, such as CenA, NGC~253 and M82, are naturally understood: lightest UHECR nuclei cannot reach us from the Virgo distance of twenty Mpc, due to their nuclei fragility above a few Mpc distances.
 Their deflection and smearing in wide hot spots is better tuned to the lighter nuclei than to the preferred proton or heavy nuclei candidate courier.
  We note that these lightest nuclei still suffer of a partial  photodistruction even from such close sources.  
  Therefore, their distruption in fragments, within few tens EeV multiplet chain of events, have been expected and later on observed by Auger collaboration, nearly a decade ago. These multiplet presences, strongly correlate with the same CenA, NGC253 sources. The statistical weight of such correlation is reminded. 
  We conclude that the  same  role of NGC~253 clustering at lower energies could also feed the Auger dipole anisotropy at lower energy ranges. 
  Such lower energy anisotropy could be fed and integrated by nearest Vela,  Crab, LMC and Cas A contributes. 
  In our present UHECR model, based on lightest nuclei in local volumes of a few Mpcs, closest AGN, Star-Burst or very close SNR are superimposing their signals, frozen in different epochs, distances and directions, feeding  small and wide anisotropy. Possible tests to confirm, or untangle the current model from alternative ones, are suggested and updated.
}
\maketitle
\section{Introduction}

Cosmic rays (CR) mainly contain charged particles.
As charges they are bent, deflected and smeared in their flight through cosmic and galactic magnetic fields losing most of the astrophysical source imprint.
Thus for a century the CR origin remained a mystery and a compelling question continually addressed by high-energy astrophysics\cite{aloisio2023ultra}.
Indeed the Ultra High Energy CR (UHECR) nature itself connects the most violent sources of the Universe with the deep inner secrets of nuclear and particle physics.

     The extreme boundaries of micro and macro physics find a rare connection by the UHECR-source correlation. We offer here our reading key based on last decades discovered maps and composition data. 
      UHECR proton above EeV ($10^{18}$ eV) are expected to be less deflected because of their harder rigidity. 
      They may more closely represent the location of their source leading to connections on an astrophysical or cosmic scale.\\
 With increasing energy the UHECR protons are deflected less and less, but above nearly 60 EeV where they should be somewhat beamed within a few degrees, they also become opaque and constrained by the Cosmic Black Body, due to the photopion production. That effect is known as the GKZ cutoff \cite{greisen1966end,GZK(1966)b}.\\  
Such UHECR nucleons are then enclosed within one to two hundred Mpc leading to a small volume bound within our much larger Universe ($4$Gpc). This proton scenario, if real, could facilitate the identification of the source associated with such an inhomogeneous scale mass distribution.

UHECRs could also be nuclei, but these are affected by increased deflection and smearing as well as partial opacity of the photopions. In addition, the lighter UHECR nuclei undergo additional photon-nuclear destruction, reducing the UHECR propagation distance to a few Mpc. Their fragments at tens of EeV, must be lighter and more diffused.
Therefore, the path of an incoming UHECR depends not only on the energy but also on the mass and composition of the charge (such as p,He,N,Fe,Ni..), on the intensity of the magnetic fields encountered and on the timing of any in-flight decays.

Indeed, for a more in-depth understanding of the phenomenon we have to consider not only the distribution of the possible sources, but also the \textit{history} of the CR, the timing of the detected signals, the delay due to the random-walk \cite{fargion2018POS}, each of them setting specific limits to the allowed energy windows.
          
Candidate sources, whether they are small as SN, GRB, Star-Burst, or larger as AGN Jet, can be found both far from us and close --even in our galaxy-- in a complex inhomogeneous cosmic structure.
We show how this difficult source correlation search can be approached and partially solved by constrains based on last two decades of observational discoveries~\cite{aloisio2023ultra} on UHECR composition, their clustering and anisotropies at different energy windows.

\section{UHECR composition and Virgo absence}
In the early 2000s most of the theoretical models were based on a proton composition for UHECR at the GZK energy limits: $6\cdot 10^{19}$-$10^{20}$ eV. In fact, the GZK cutoff was the main focus looked for in the UHECR spectra. The proton at such energies should cluster within tight angles, $2^{\circ}$ for coherent bending, or near $8^{\circ}$ for incoherent random bending \cite{fargion2011coherent}; it must also be contained in volumes of $1-2$ hundred Mpc, the GZK volume. %
Thus they were expected to rise and shine from the nearest and more mass populated regions, similar to how infrared galaxies are painting the infrared sky: the Virgo cluster,  at a distance of about 20 Mpc, is the dominant mass in the Auger southern emisphere sky. 
Surprisingly, at those energies the UHECR signals in Auger, and later also in Telescope Array (TA) %
in the northern hemisphere, did not show a sharp signature %
confirming this expectation: the \textit{surprising Virgo absence}.

     Rather both Auger and TA discovered some wider ($8^{\circ}-20^{\circ}$) clustering elsewhere, uncorrelated with Virgo. %
      A first Hot Spot in southern sky was pointing to Cen-A, the nearest AGN in Auger sky. A second Hot Spot in northern sky was found by TA pointing toward a nearest AGN M82. A third clustering, or Hot Spot, was later recognized \cite{fargion2015meaning00}, pointing to NGC~253 a micro AGN or star-burst galaxy still in the Auger sky.

\begin{figure}
\begin{center}
\includegraphics[width=.95\columnwidth]{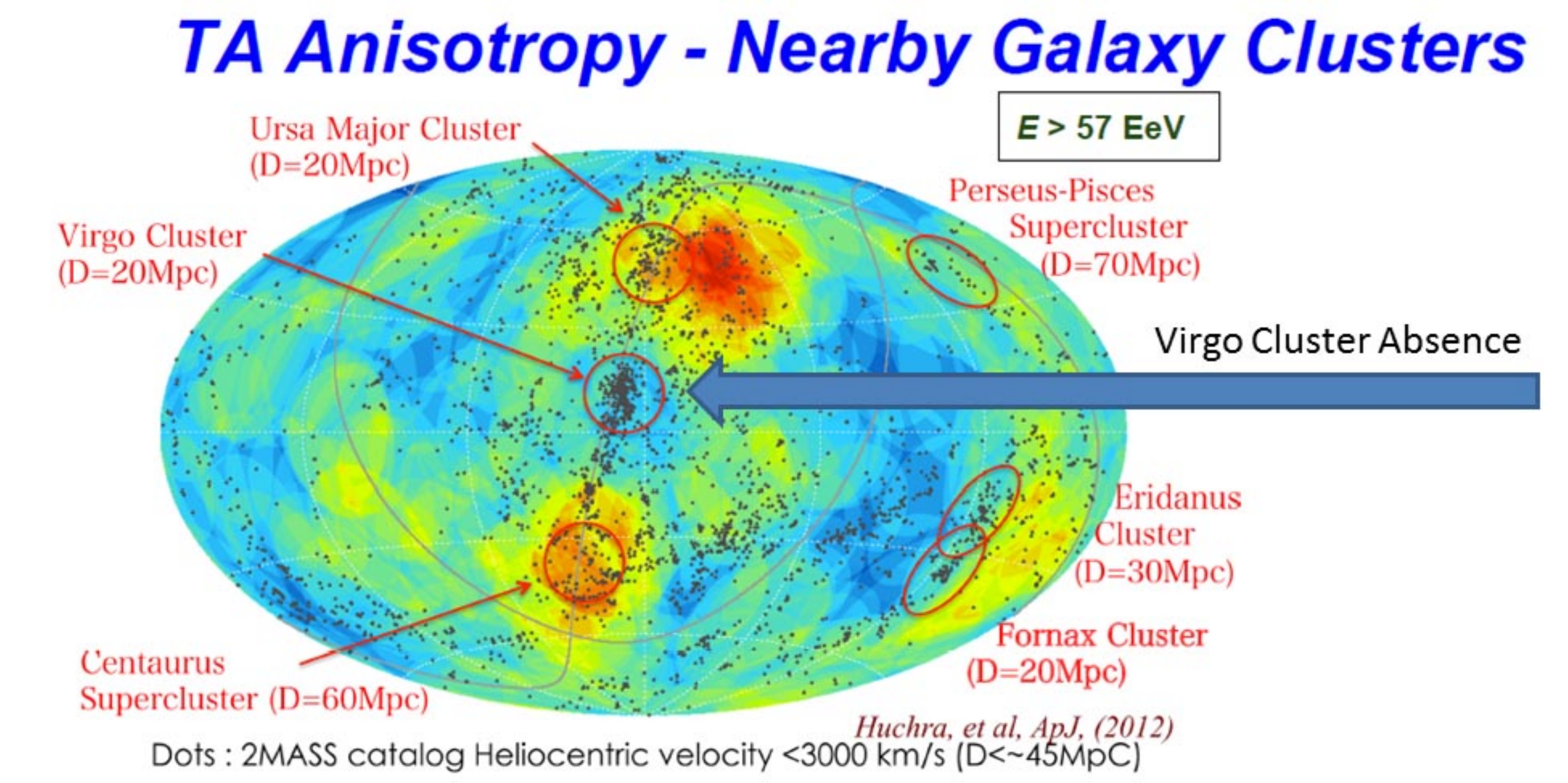}
\caption{
The combined  map, in equatorial coordinates, based on the  Auger and TA data. Note in red two Hot Spot UHECR clustering signals. The highest UHECR density event rates are north around M82 AGN for TA and south around Cen-A for Auger.
    The dark dots circled in red in the center \cite{Huchra2012RedShift} correspond to the closest potential sources included in the GZK volume, pointing to the Virgo cluster.
    Those sources were ideal for any UHECR proton carrier. The lighter nuclei are much more limited, a few Mpc, by photonuclear destruction. These are the ideal candidates who can explain the absence of Virgo in the Auger map.
}
\label{Fargionfig1}
\end{center}
\end{figure}

    A combined map of the Auger and TA signals is presented in equatorial coordinates in Fig.\ref{Fargionfig1}.
In the center the area with the most abundant dark dots, thousands of galaxies of the Virgo Cluster, at a distance of 18 Mpc well within the GZK volume~\cite{Huchra2012RedShift}.   
  
Assuming UHECR protons, the Virgo absence was (and still is) puzzling and inexplicable. Since 2007 a harder spectra in Auger composition studies \cite{watson2023model} based on the recently observed most energetic UHECR airshower profile, disfavored protons above 10~EeV in favour of light and lightest nuclei. Protons have actually only been found to be dominant at sub-EeV energy ranges.  %
This led us to suggest that the absence of Virgo could be explained if UHECRs consist mainly of lighter nuclei, such as D, He, Li, Be ~\cite{fargion2008light,fargion2012cen,fargion2010uhecr,FARGION2009162}.  
     
The lightest nuclei are indeed too fragile to survive the 20 Mpc flight from Virgo, due to photon-nuclear distruption. On the other hand, Cen-A, M82 and NGC~253 are just few Mpc away, much closer than Virgo. 
These smaller distances allow most UHECRs to survive photon nuclear distruption and opacity. These sources are among the ideal candidates star-burst or active AGN around us, whose signals are well tuned and observable in the Auger-TA data. As the data of the last decade had shown.
Only a decade later (2017), the UHECR signature of the light nuclei was largely confirmed (see note on page 26 of \cite{PAO(2017b)}) by their detailed slant depth averaged models of the UHECR shower maximum, which show the key role of light and even the lightest nuclei.

Therefore,  around $3\cdot 10^{19}$ eV, UHECRs mainly He like nuclei  may be cosmic but, as already mentioned,  extremely  local just within a few Mpc, almost all in our Local Group \cite{fargion2015meaning00}. 

          Other near and far cosmic UHECR sources could further add dilute homogeneous and isotropic noises. Rarer and heavier nuclei, such as Ni and Fe, may also be present at higher energies (around or higher than $7\cdot10^{19}$ eV), but so bent and smeared that they are often confined within our own galaxy \cite{fargion2010uhecr}. Ultimately these near candidates, Cen-A, M82, NGC~253, are contributing the most to the local anisotropy~\cite{fargion2015meaning00}.

 \subsection{Hot Spot and correlated multiplet fragments}
  A lightest nuclei whose charge is two or four times that of the proton undergoes a somewhat larger deflection than the protons, a smearing comparable to those observed in the Southern and Northern skies.  
 The charge $Z=2-4$ of lightest nuclei explains the observed sizes of the Hot Spots due to their random incoherent deflection at about $3 \cdot 10^{19}$ eV, assuming He-like nuclei are bent within angles of $16^\circ-32^\circ$ \cite{fargion2011coherent}.

      We note that  other light but heavier nuclei, like N,O, suffer no photo-nuclear distruption from Virgo, but are more smeared in a wider solid angle.    
These nuclei would cause a signal from Virgo diluted in a large angle of the sky, $56^\circ-64^\circ$, with a possible negligible role in the observed Hot Spot, but with a potential role in the dipole anisotropy. Such a dipole signature toward Virgo is again absent: it points elsewhere, almost half the sky away.
Therefore, this is a compelling argument for considering UHECR nuclei consisting mostly of the lightest ones, unable to reach us from Virgo.

    The same photo-nuclear distruption can occur for D, He,Li, Be nuclei even from a nearby Cen-A, or NGC~253; in part it had to happen, and as we have noted it happened. These fragments were to be centered or distributed  along the main Hot Spot sources in the Auger sky: Cen-A, as well as in the  additional clustering around NGC~253. M82 is not visible from Auger sky, but is within the TA sky. 
    
\begin{figure}
\begin{center}
\centering\includegraphics[width=0.99\columnwidth]{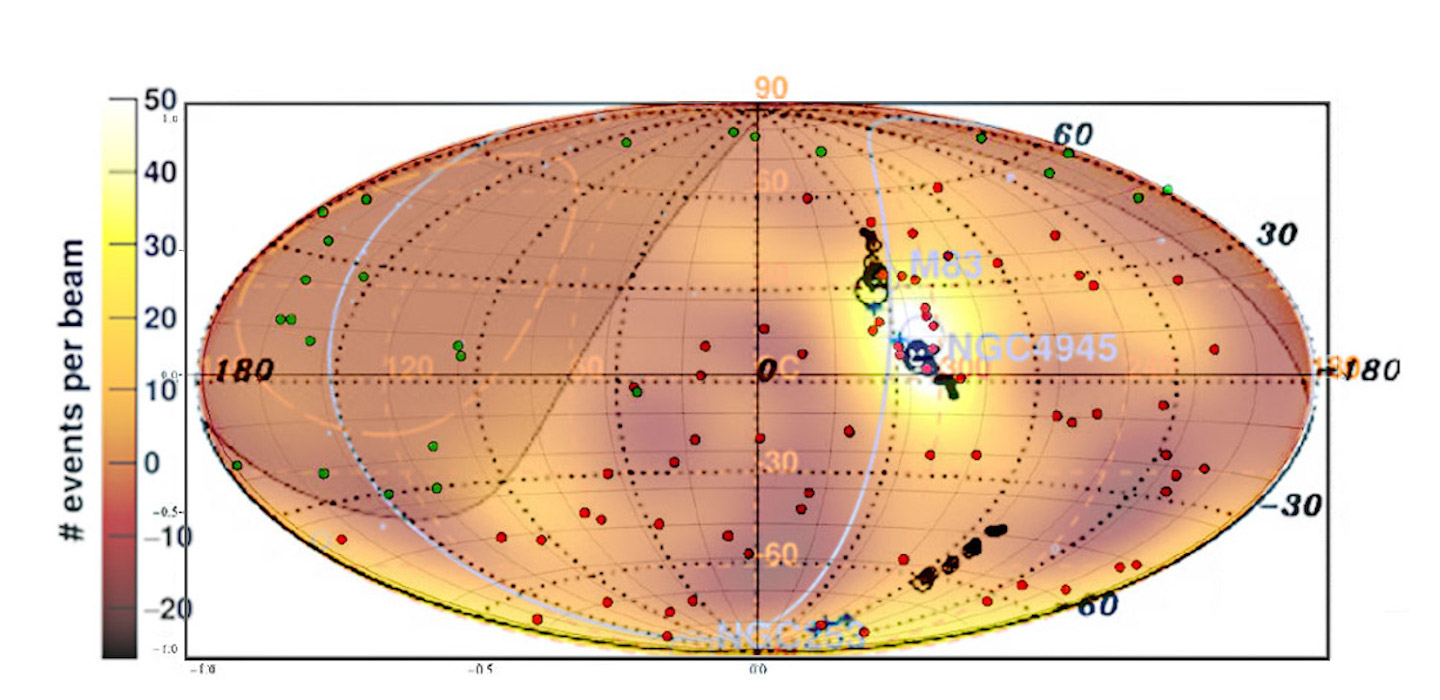}
\caption{%
The galactic UHECR map overlaid on the recent $39$ EeV anisotropy map, shows the earliest (2011) UHECR events and the associated 20 EeV smeared multiplet, up and down respect to the Cen~A position. 
We interpret these nearly vertical spread of events, followed by the fragment signals, as due to the horizontal planar galactic spiral magnetic fields, whose spin flip up and (later) down during the UHECR flight, is leading to an up-down deflection of the charged He,as well  as D, Li, Be, lightest UHECR fragments.
 Let us note, here for the first time, also the presence of a train of UHECR fragments located around the SMC and LMC galactic direction, pointing to NGC~253 source in the South galactic Pole \cite{abreu2012search}.  The probability that these fragments are correlated by chance  with both to the two main candidate of UHECR in Auger sky, Cen-A and NGC~253, is quite negligible, below $0.001\%$.}
\label{Fargionfig6}
\end{center}
\end{figure}

    We had foreseen \cite{fargion2010uhecr,fargion2011coherent}, such a tail of fragments around Cen-A, a couple of years before its observation by Auger \cite{abreu2012search} in 2012, see Figure \ref{Fargionfig6}.
In the figure three multiplets of events are marked with black dots. Their extrapolated endpoints are signed with a blue cross. We note that two of them are within $8^\circ$ of Cen-A. And the third one is 
within $8^\circ$, to the other relevant source, NGC~253.
    
      One may inquire how often such a correlation may occur by chance.
       The solid angle distance of each source from its fragment tail cross center is about $7^\circ - 8^\circ$, which represents less than $1\%$ of the AUGER sky.

      The probability that two such related events occurred within 3 trials, each of which point to the same Cen-A area within $1\%$ of the sky, is less than $2.97 \cdot 10^{-4 }$. 
      
      Furthermore, the probability that three signals appearing above the two selected sources, in an area that is $2\%$ of the Auger sky, and occurring by chance is about $8 \cdot 10^{ -6} $. We recall that Cen-A and NGC~253 are the only two in the Auger sky.
      
      Thus, the directional clustering of the fragments towards our a priori selected sources offers another strong argument in favor of the current model of lighter nuclei. %
      
\begin{figure}
\begin{center}
 \centering\includegraphics[width=0.90\columnwidth]{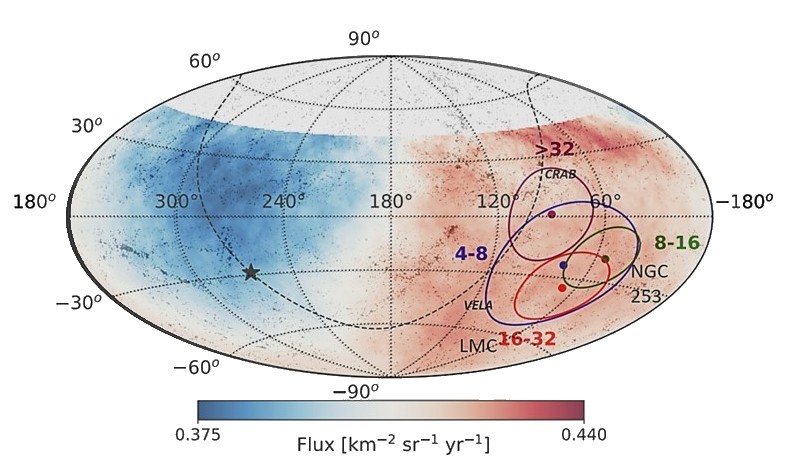}
\caption{%
The UHECR dipole clustering in equatorial coordinates. Blue, green, red, and violet points and curves indicate different energy ranges (4-8, 8-16, 16-32 and $>$32). Few suggested galactic or nearby extragalactic source candidate are labeled. 
The main extragalactic source, the star-burst NGC~253 role, is well correlated to the $8-16; 16-32$ EeV anisotropy energy ranges. An additional minor influence of galactic sources, as Crab, Vela and LMC, could explain the lower energy, $4$ EeV, anisotropy and its directional variability with the different energy increase.
}
\label{dipole2c}
\end{center}
\end{figure}

\begin{figure}[t]
\begin{center}
 \centering\includegraphics[width=0.90\columnwidth]{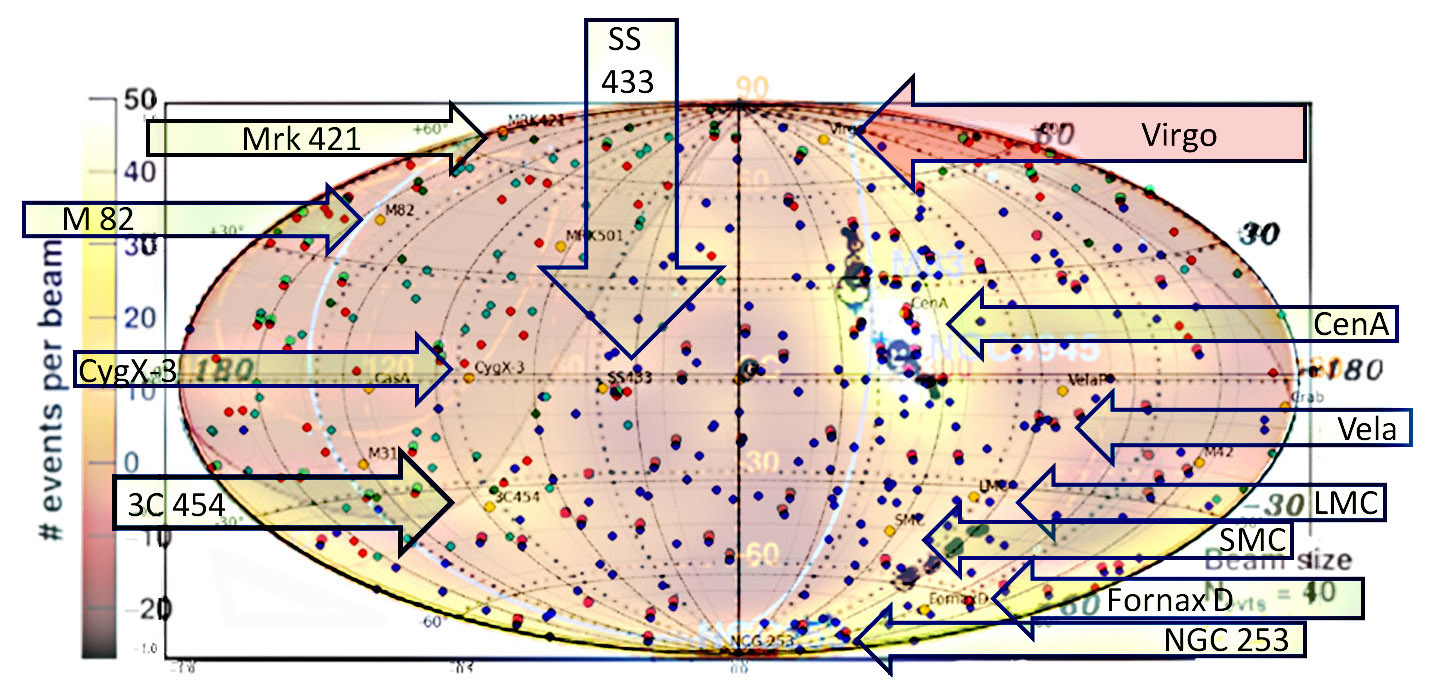}
\caption{%
UHECR map in galactic coordinate with marked the main probable identified sources and the missing ones as Virgo.
The role of  3C~454 may be relevant only if the narrow clustering will be more enhanced~\cite{Fargion2015Meaning}. 
A $39$ EeV anisotropy and a $20$ EeV multiplet map is also overlaid. 
These candidate sources are the same ones discussed since 2015 \cite{fargion2015meaning00}. 
In particular we note the  growing signals around the Vela, LMC, SMC and Crab in nearby sky, possibly feeding the low energy dipole Auger anisotropy. The extreme powerful gamma source, 3C~454, shows some minor clustering of UHECR. The 3C~454 is too far in GZK distance unity to be explained via any directional nuclear event. It  can be an UHECR secondary source: an eventual UHE ZeV neutrino courier, whose hitting the relic cosmic neutrino background, in a hot wide neutrino halo~\cite{Fargion.et.al.(1999)} is leading to the Z boson capable to decay also into UHECR nucleon or anti-nucleon. The relic neutrinos are possibly, also,  at $\simeq 1.6 eV$ mass, as the recent candidate sterile ones. The ZeV scatterinng  could be leading to Z resonant bosons. The secondary decay of such ultrarelativistic Z boson, may shine as a final UHECR nucleon and antinucleon, observed finally on Earth as a minor UHECR clustering along the 3C~454 source \cite{fargion2018POS}.
}
\label{Fargionfig8}
\end{center}
\end{figure}

  \subsection{UHECR at AGN 3C 454: a Z resonance?}
Few unexplained UHECR events correlated with very distant AGN (such as 3C~454), if confirmed, may require a different mechanism to exceed the GZK cut-off. A ZeV neutrino that scatters on the relic neutrinos, forming the final UHECR and effectively exceeding any GZK cut-off.
    The  calorimeter for ZeV neutrinos scattering is in the huge light (even sterile) relic cosmic neutrinos halo spread around origin galaxy or local group. 
They are converting the ultrarelativistic energy of the Z boson, by its decay, into secondary nucleons and antinucleons. These nucleons can reach us, overcoming any GZK distance. These UHE ZeV neutrinos \cite{Fargion.et.al.(1999)}, complementary to the GZK neutrinos at tens of EeV, could rise as ascending or horizontal airshowers in deep valleys \cite{Fargion(2000)}, induced by a tau decay skimming and escaping from the Earth \cite{Fargion.et.al.(2004)}.
  
   \section{Conclusion}
  Many scenario inspired by many spare parameter are confusing the UHECR connection to their sources. The UHECR  anisotropy \cite{Mayotte2022Aq} and asymmetry \cite{plotko2022indication} show that their main origin is within a very local Universe, not a hundred Mpc away.
  
    UHECR are very probably formed in jets both in  AGN quasars and in micro-quasars (star forming regions, accretion disk and ejecta), powered by black Hole or neutron star binary tidal disruptions. These jets \cite{fargion1999nature} may arise in both AGN and in smaller size GRB, micro quasars and SGR, shining at different epochs and locations. 
    The charged hadronic nuclei UHECR flying from those jets are possibly bent, spread by random walk, suffering large time delay. 
    This explain why they are not expected to be correlated with present optical active AGN at distances of hundred Mpc\cite{fargion2018POS}. 
There is a possibility that the heavier UHECRs such as NI, Co, Fe are mostly bent, twisted and even contained in spiral paths within halos of their own origin.
    The  lightest nuclei \textit{Helium-like}, possibly  secondary fragments of heavier ones,  are quite directional; they are able to escape from those AGN galaxy, star-burst sources as nearby Cen-A, NGC~253, M82,  reaching us with some memory of their origin in  the observed few Hot Spots. 
    
These lightest UHECR nuclei at tens of EeV have the virtue of flying only within a space of very few Mpc, hiding the presence of Virgo and offering only a local group detection view.
Protons only play a role at EeV or lower, escaping hundreds of Mpcs rather smeared in their arrival directions.
Their bending and overlapping should lead to a homogeneous isotropic sky, like the one observed at the EeVs energy windows. Lastly, the very few and rare heaviest and more energetic Ni-Co-Fe UHECRs can also arise around our galactic halo, perhaps also polluted by nearest galactic sources.
    
    In summary lightest nuclei in UHECR are tagging just a few sources in the very local universe Cen-A, M82, NGC~253; AGN or starburst or microquasars.
   Also at lower  energy, in the $4-30$ EeV range, UHECR are feeding  an Auger dipole anisotropy, mostly linked to NGC~253, see Fig.\ref{dipole2c}, and partially to Vela, Crab, LMC, SMC. 
   Other minor, unexplained UHECR cluster can, if confirmed,  revive earliest Z boson resonance models, based on UHE ZeV neutrinos scattering  on relic cosmic ones, with mass, see Fig.\ref{Fargionfig8}.
   The eventual multi-peak energy signature \cite{Fargion2007292} in their spectra, may in principle in a far future, test the different neutrino mass splitting or even the recent sterile neutrino masses candidature.   
   Their final nucleon anti-nucleon shower composition signature (not the lightest nuclei one) is a  key test, to verify and disentangle their Z boson nature, see Fig.~\ref{Fargionfig8}, \cite{fargion2018POS}.
   
   In conclusion, early popular proton courier at the GZK energy of $40-60$ EeV, is not longer a viable  UHECR candidate \cite{watson2023model}. 
   The recent,  simplest models based on  star-burst versus AGN nearby sources nature, while ignoring the UHECR nuclei composition role, cannot explain the puzzling absence of Virgo.   
   In a sentence, for the moment, lightest nuclei \cite{fargion2008light} and their clustering along a few correlated local sources \cite{fargion2015meaning00} are the first and main guaranteed messages from last decade of Auger and TA data understanding.

The three tens of EeV  events in the Auger sky clustered in multiplets  pointing two towards Cen-A and one towards NGC~253, as predicted \cite{fargion2012tev,fargion2012cen,fargion2010uhecr} and observed \cite{abreu2012search}  (see Fig.~\ref{Fargionfig6}), are providing key support for the lightest nuclei model.

\subsection*{Acknowledgement}
\vspace{-0.3cm}
{\footnotesize  \addtolength{\itemsep}{-0.1in}
The research by M.K. was financially supported by Southern Federal University in the framework of the State contract with the Ministry of Science and Education of Russian Federation.
}

\FloatBarrier   

\bibliography{daf2022}

\end{document}